\documentclass[a4paper,11pt]{article}
\pdfoutput=1
\usepackage{jcappub}
\usepackage{slashed}
\usepackage{mathtools}
\usepackage[T1]{fontenc}
\usepackage[normalem]{ulem}

\graphicspath{{plots/}}

\newcommand{\gs}{g_\star}
\newcommand{\gss}{g_{\star s}}
\newcommand{\Tbbn}{T_\text{BBN}}
\newcommand{\Tmax}{T_\text{max}}
\newcommand{\Trh}{T_\text{rh}}
\newcommand{\Hrh}{H_\text{rh}}
\newcommand{\arh}{a_\text{rh}}
\newcommand{\sv}{\langle\sigma v\rangle}

\title{Probing low-reheating scenarios with minimal freeze-in dark matter}
\author[a]{Nicolás Bernal,}
\author[b]{Chee Sheng Fong,}
\author[c]{\'Oscar Zapata}

\affiliation[a]{New York University Abu Dhabi\\
PO Box 129188, Saadiyat Island, Abu Dhabi, United Arab Emirates}
\affiliation[b]{Centro de Ciências Naturais e Humanas,
Universidade Federal do ABC\\
09.210-170, Santo André, SP, Brazil}
\affiliation[c]{Instituto de Física, Universidad de Antioquia\\
Calle 70 \# 52-21, Apartado Aéreo 1226, Medellín, Colombia}

\emailAdd{nicolas.bernal@nyu.edu}
\emailAdd{sheng.fong@ufabc.edu.br}
\emailAdd{oalberto.zapata@udea.edu.co}

\abstract{The parameter space of freeze-in dark matter (DM) with mass $m_\chi$ through light dark photon (``minimal freeze-in DM'') is currently being probed by direct detection experiments through electron and nuclear recoil. Exploring the DM production in the mass range $10^{-2}~{\rm MeV} < m_\chi < 10^3$~TeV, we quantify the impact of quantum statistics and the reheating dynamics (beyond the instantaneous reheating approximation) on the DM production in the early universe, in particular, the dependence on the cosmic equation of state and the scaling of the temperature of the Standard Model bath during reheating. Special cases corresponding to matter-domination and kination are carefully studied. To fit the entire observed DM relic abundance, low-temperature reheating scenarios require an increase in the coupling between dark and visible sectors which, in turn, enhances the regions of the parameter space that are already tested and will be probed by next-generation direct detection experiments for diverse reheating scenarios.}
\begin{document}
\begin{flushright}
\end{flushright}
\maketitle

%%%%%%%%%%%%%%%%%%%%%%%%%%%%%%%%%%%%%
\section{Introduction}
%%%%%%%%%%%%%%%%%%%%%%%%%%%%%%%%%%%%%
To date, the nature of dark matter (DM) remains an open question, despite being about five times more abundant than visible matter~\cite{Planck:2018vyg}.  One intriguing solution to this puzzle is the feebly-interacting massive particle (FIMP) paradigm. Unlike popular weakly-interaction massive particles (WIMPs)~\cite{Arcadi:2017kky, Roszkowski:2017nbc, Arcadi:2024ukq}, FIMPs were produced in the early Universe through the freeze-in mechanism~\cite{McDonald:2001vt, Choi:2005vq, Kusenko:2006rh, Petraki:2007gq, Hall:2009bx, Elahi:2014fsa, Bernal:2017kxu}, whereby their extremely weak interactions with the standard model (SM) plasma kept them from reaching thermal equilibrium.

Although the extremely weak interactions of FIMPs suggest that testing the freeze-in mechanism might be highly challenging, recent developments have shown this to be increasingly feasible. For example, an increase in DM direct detection rates occurs in the case of GeV DM with a light mediator~\cite{Hambye:2018dpi}, in cases with sub-GeV DM particles~\cite{XENON:2019gfn, XENON:2020rca, Elor:2021swj}, or due to the boost induced by the scattering of DM particles off electrons in the Sun~\cite{An:2021qdl, CDEX:2023wfz, Emken:2024nox}. In addition, the feeble coupling between the DM and the mediator can result in a sizable decay length for the mediator, producing distinctive signatures in colliders that can be observed, such as displaced vertices or long-lived particles~\cite{Molinaro:2014lfa, Hessler:2016kwm, Co:2015pka, Belanger:2018sti, Junius:2019dci, Calibbi:2021fld, Barman:2024nhr, Barman:2024tjt}.

In this work, we consider one of the simplest models of DM with a very light vector mediator: a massive dark photon $\gamma'$ with mass $m_{\gamma'}$ that couples to the fermionic DM $\chi$ and $\overline{\chi}$ of mass $m_\chi$ with dark charge $e'$, and to electrically charged particles of the standard model (SM) with an effective charge $\epsilon\, e$ suppressed by $\epsilon \ll 1$, where $e \sim 0.3$ is the electric charge~\cite{Ackerman:2008kmp, Feng:2008mu, Feng:2009mn, Chu:2011be}. In this minimal scenario, DM is produced through the freeze-in paradigm from the scattering of SM particles $\psi$ to a pair of DM $\psi \overline\psi \to \chi \overline\chi$ and plasmon decays $\gamma^* \to\chi\overline{\chi}$, which becomes important for $m_\chi \lesssim m_e$ when the electron number density starts to be Boltzmann suppressed, making $e^+ e^- \to \chi \overline\chi$ less efficient~\cite{Dvorkin:2019zdi}. Since the cross section of freeze-in scattering is proportional to $\epsilon^2\, e^2\, {e'}^2/T^2$, most of the DM production occurs at low SM temperatures $T$, when $T \sim m_\chi$; a typical behavior of an infrared-dominated FIMP production.

It is important to keep in mind that, in addition to the particle-physics framework, the cosmological background on which DM is produced has to be under control. In the simplest scenario, DM could have been generated during a phase in which the expansion of the universe was dominated by SM radiation and the SM entropy was conserved: that is, the standard cosmological scenario. However, deviations from this minimal paradigm could also have occurred before the Big Bang nucleosynthesis epoch~\cite{Allahverdi:2020bys, Batell:2024dsi}. For example, after cosmic inflation, the universe has to transit from an inflaton domination to a SM radiation domination: the cosmic reheating phase~\cite{Kofman:1997yn}. During reheating, the maximum SM temperature reached by the plasma could be much higher than the reheating temperature $\Trh$~\cite{Giudice:2000ex}. Additionally, the evolution of the background (i.e. equation of state of the inflaton and evolution of the SM temperature) is uncertain.

Then, it becomes clear that the DM production in the early universe is highly-impacted by the evolution of the background. In particular, if the reheating temperature $\Trh$ is smaller than $m_\chi$, here denoted ``low reheating'' scenarios, the production would be Boltzmann suppressed due to kinematics and therefore the coupling required $\epsilon\, e'$ has to be enhanced accordingly~\cite{Boddy:2024vgt,Gan:2023jbs}. We will show that when the reheating process is prolonged (depending on the coupling of inflaton fields to radiations), the freeze-in production is sensitive to the reheating dynamics, particularly to the cosmic equation of state $\omega$ during that period. Due to the scaling of $1/T^2$ in the scattering cross section, the production is not sensitive to the highest temperature during reheating $\Tmax$ and this allows us to determine a finite region of parameter space for $\Tmax \gg \Trh$ and $\Tmax = \Trh$, the latter corresponds to an instantaneous reheating scenario of Ref.~\cite{Boddy:2024vgt}. Finally, this model has become the benchmark scenario for the freeze-in model due to the enhanced scattering cross section for direct detection which scales as $1/v^4$ where $v \sim 10^{-3}$ is the relative velocity between the DM and the detector target (electron or nucleus)~\cite{Essig:2011nj, Essig:2012yx, Essig:2015cda} and therefore allows low reheating scenarios to be probed.\footnote{A complimentary approach is to use particle physics experiments (collider and/or beam dump) to produce and detect ``millicharged DM'' and thereby constrain the corresponding reheating temperature in order not to overclose the universe~\cite{Gan:2023jbs}. However, bounds from DM direct detection are much stronger~\cite{Iles:2024zka}.}

The paper is organized as follows. In the next section, the particle physics and the cosmological setups are introduced. In Section~\ref{sec:DM}, we present a comprehensive analysis of the parameter space for DM production in low-reheating scenarios. We identify regions that are consistent with all theoretical and observational constraints, as well as regions that are within the sensitivity of current and future experiments. Finally, we conclude in Section~\ref{sec:concl}. In Appendix~\ref{app:quantum_statistics}, we describe the quantum statistical effects associated with the freeze-in mechanism.

%%%%%%%%%%%%%%%%%%%%%%%%%%%%%%%%%%%%%
\section{The Set-up} \label{sec:framework}
%%%%%%%%%%%%%%%%%%%%%%%%%%%%%%%%%%%%%
%%%%%%%%%%%%%%%%%%%%%%%%%%%%%%%%%%%%%%%%
\subsection{Particle Physics}
%%%%%%%%%%%%%%%%%%%%%%%%%%%%%%%%%%%%%%%%
The model we consider consists of a $U(1)'$ gauge extension,  with a corresponding very light gauge boson $\hat{X}_\mu$, and a Dirac fermion $\chi$ with a $U(1)'$ charge $e'$, being a singlet of the SM gauge group. The SM matter content does not transform under $U(1)'$; the hidden and visible sectors are connected through the kinetic mixing term involving the SM hypercharge gauge boson $\hat{A}_\mu$. The new terms in the vacuum Lagrangian density can be written as~\cite{Hall:2009bx, Chu:2011be, Essig:2011nj}
\begin{equation} \label{eq:Lag}
    \mathcal{L}_D=\, -\frac14\, \hat{X}_{\mu\nu}\, \hat{X}^{\mu\nu} + \frac12\, m_{\gamma'}^2\hat{X}_\mu\, \hat{X}^\mu + \frac{\epsilon_Y}{2}\, \hat{X}_{\mu\nu}\, \hat{B}^{\mu\nu} + \overline{\chi}\, (i\slashed{\partial} - m_\chi)\, \chi - e'\, \hat{X}_\mu\, \overline{\chi}\, \gamma^\mu\chi\,,
\end{equation}
where $\hat{X}_{\mu\nu} \equiv \partial_\mu \hat{X}_\nu - \partial_\nu \hat{X}_\mu$ and $\hat{B}_{\mu\nu}\equiv \partial_\mu \hat{A}_\nu - \partial_\nu \hat{A}_\mu$; and where four new free parameters $m_{\gamma'}$, $\epsilon_Y$, $m_\chi$, and $e'$ were introduced. After the electroweak symmetry breaking, the field redefinition~\cite{Fabbrichesi:2020wbt, Bhattiprolu:2024dmh}
\begin{align}\label{eq:field-rot1}
    \hat{A}_\mu &= A_\mu+\epsilon A'_\mu\,,\\\label{eq:field-rot2}
    \hat{X}_\mu &= A'_\mu-\epsilon\,\tan\theta_W\,Z_\mu\,,\\\label{eq:field-rot3}
    \hat{Z}_\mu &= Z_\mu\,,
\end{align}
with $\epsilon \equiv \epsilon_Y\cos\theta_W$, allows to go from the gauge basis to the mass basis. Therefore, Eq.~\eqref{eq:Lag} can be rewritten as 
\begin{equation} \label{eq:original_mass_basis}
    \mathcal{L}_D \supset\, \frac12\, m_{\gamma'}^2\, A'_\mu\, {A'}^\mu -\epsilon\, e\, A'_\mu\, J^\mu_{\rm EM} -  e' \left(A'_\mu - \epsilon\, \tan\theta_W\, Z_\mu\right) \overline{\chi}\, \gamma^\mu\, \chi\,,
\end{equation}
with canonical kinetic terms for both SM and dark photons, and where $J_{\rm EM}^\mu$ denotes the electromagnetic current. The effective portal coupling between the two sectors can be expressed as~\cite{Chu:2011be}
\begin{equation}
    \kappa \equiv \frac{\epsilon\, e'}{e}\,.
\end{equation}

It is important to note that, due to interactions with charged particles in the thermal plasma,  SM photons gain a thermal mass $m_\gamma$ of the order of the SM temperature $T$. To see the consequences of this term, let us introduce the following in-medium mass terms to Eq.~\eqref{eq:original_mass_basis}~\cite{Dvorkin:2019zdi}
\begin{equation}
    {\cal L}^{\rm IM}_{\rm mass} = \frac12\, m_{\gamma}^{2}\, A_{\mu}\, A^{\mu} + \epsilon\,  m_{\gamma}^{2}\, A'_{\mu}\, A^{\mu},
\end{equation}
and after rotating to the effective mass basis photon $\tilde{A}_{\mu}$ and dark photon $\tilde{A}'_{\mu}$, we have that the in-medium Lagrangian up to $\mathcal{O}(\epsilon)$ takes the form~\cite{Knapen:2017xzo, Dvorkin:2019zdi}
\begin{align} \label{eq:effective_mass_basis}
    {\cal L}^{\rm IM}_D &\supset \frac12\, m_{\gamma'}^{2}\, \tilde{A}'_{\mu}\, \tilde{A}'{}^{\mu} + \frac12\, m_{\gamma}^{2}\, \tilde{A}_{\mu}\, \tilde{A}^{\mu} + e \left(\tilde{A}_{\mu} + \frac{\epsilon\, m_{\gamma'}^{2}}{m_{\gamma'}^{2} - m_{\gamma}^{2}}\, \tilde{A}'_{\mu}\right) J_{{\rm EM}}^{\mu}\nonumber \\
    &\qquad + e' \left(\tilde{A}'_{\mu} - \frac{\epsilon\, m_{\gamma}^{2}}{m_{\gamma'}^{2} - m_{\gamma}^{2}}\, \tilde{A}{}_{\mu} - \epsilon\, \tan\theta_{W}\, Z_{\mu}\right) \overline{\chi}\, \gamma^{\mu}\, \chi\,.
\end{align}

In the setup considered in this work, we consider $m_{\gamma'}$ to be sufficiently small and we remain agnostic about the mechanism through which the dark photon acquires mass, whether through the Stückelberg~\cite{Stueckelberg:1938hvi} or the Brout-Englert-Higgs mechanisms~\cite{Englert:1964et, Higgs:1964pj}. For definiteness, we will restrict ourselves to $m_{\gamma'} \lesssim 10^{-21}$~MeV to avoid strong constraints ($\epsilon \lesssim 10^{-7} - 10^{-5}$) due to resonant conversion $\gamma \to \gamma'$ that can distort the cosmic microwave background~\cite{Caputo:2020bdy, Garcia:2020qrp, Witte:2020rvb}; see Ref.~\cite{Fabbrichesi:2020wbt} for other constraints in the case of larger values of $m_{\gamma'}$. With $m_{\gamma'}\lesssim10^{-21}$~MeV and $m_{\gamma}\sim 0.1\, T$, the photon thermal mass always dominates $m_{\gamma}\gg m_{\gamma'}$ in the early universe or in stellar environments, resulting in a suppressed coupling of $J_{{\rm EM}}^{\mu}$ to the dark photon by $\epsilon\, (m_{\gamma'}/m_{\gamma})^2$. Therefore, the dark-photon production in the early universe is negligible~\cite{Dvorkin:2019zdi} and the stellar constraints on $\epsilon$ from the emission of Stückelberg dark photons is also correspondingly relaxed~\cite{An:2013yfc, An:2013yua}. 

Notice that Eqs.~\eqref{eq:original_mass_basis} and~\eqref{eq:effective_mass_basis} both give the same effective coupling $\kappa\, e^2$ in the amplitude between the SM charged particles and the DM and hence, for the very light $m_{\gamma'} \ll \mathcal{O}(1)$~eV that we consider here, the freeze-in production of DM from the thermalized SM sector becomes equivalent to the one in the massless dark photon scenario discussed in Ref.~\cite{Chu:2011be}. Furthermore, the ``millicharged DM'' bounds also apply, and as will be seen in Section~\ref{sec:DM}, for $m_\chi \lesssim 0.1$ MeV, there are relevant bounds on $\kappa$ from stellar energy loss due to emission of DM~\cite{Davidson:2000hf, Vogel:2013raa, Fung:2023euv}. All in all, such a small $m_{\gamma'}$ does not play a direct role in our analysis and therefore the phenomenologically relevant parameters are $m_\chi$ and $\kappa$; we will leave the investigation of larger $m_{\gamma'}$ for future work.\footnote{Refs.~\cite{Chu:2011be, Hambye:2019dwd} analyzed the production of DM relic abundance through various regimes (in addition to freeze-in), which are influenced by the chosen values of the model's free parameters: $m_\chi$, $m_{\gamma'}$, $e'$ and $\epsilon$.}

%%%%%%%%%%%%%%%%%%%%%%%%%%%%%%%%%%%%%%%%
\subsection{Cosmology}
%%%%%%%%%%%%%%%%%%%%%%%%%%%%%%%%%%%%%%%%
During cosmic reheating, the behavior of the background is uncertain. Here we parameterize the evolution of the SM temperature $T$ as a function of the cosmic scale factor $a$ as~\cite{Bernal:2024yhu}
\begin{equation} \label{eq:T}
    T(a) = \Trh \times
    \begin{dcases}
        \left(\frac{\arh}{a}\right)^\alpha &\text{ for } a_I \leq a \leq \arh\,,\\
        \left(\frac{\gss(\Trh)}{\gss(T)}\right)^\frac13 \frac{\arh}{a} &\text{ for } \arh \leq a\,,
    \end{dcases}
\end{equation}
where $a = a_I$ and $a = \arh$ correspond to the scale factors at the beginning of reheating (that is, the end of inflation) and the end of reheating (i.e. the onset of the radiation-dominated era), respectively. Furthermore, $\Trh$ denotes the SM temperature $a = \arh$. The reheating temperature $\Trh$ (that is, the SM temperature from which the Universe begins to be dominated by SM radiation) must satisfy $\Trh > \Tbbn \simeq 4$~MeV~\cite{Sarkar:1995dd, Kawasaki:2000en, Hannestad:2004px, DeBernardis:2008zz, deSalas:2015glj}, in order not to spoil the success of BBN. The function $\gss(T)$ corresponds to the numbers of relativistic degrees of freedom contributing to the SM entropy density $s$ given by
\begin{equation}
    s(T) = \frac{2 \pi^2}{45}\, \gss\, T^3.
\end{equation}
Interestingly, for $\alpha > 0$, at the beginning of reheating, the thermal plasma reaches a temperature $\Tmax \equiv T(a_I) > \Trh$~\cite{Giudice:2000ex}. After reheating (when $a > \arh$), $T(a) \propto 1/a$ as expected in an era where the SM entropy is conserved.

In the early Universe, the Hubble expansion rate $H$ has a contribution from the inflaton and SM radiation energy densities ($\rho_\phi$ and $\rho_R$, respectively), and is given by
\begin{equation}
    H^2 = \frac{\rho_\phi + \rho_R}{3\, M_P^2}\,,
\end{equation}
with $M_P \simeq 2.4 \times 10^{18}$~GeV being the reduced Planck mass, and
\begin{equation}
    \rho_R(T) = \frac{\pi^2}{30}\, \gs\, T^4,
\end{equation}
where $\gs(T)$ corresponds to the numbers of relativistic degrees of freedom contributing to the SM energy density~\cite{Drees:2015exa}. Assuming that during reheating the inflaton has an effective equation of state $\omega$, $\rho_\phi(a) \propto a^{-3(1+\omega)}$, and therefore~\cite{Bernal:2024yhu}
\begin{equation} \label{eq:H}
    H(a) = \Hrh \times
    \begin{dcases}
        \left(\frac{\arh}{a}\right)^\frac{3(1+\omega)}{2} &\text{ for } a_I \leq a \leq \arh\,,\\
        \left(\frac{\gs(T)}{\gs(\Trh)}\right)^\frac12 \left(\frac{\gss(\Trh)}{\gss(T)}\right)^\frac23 \left(\frac{\arh}{a}\right)^2 &\text{ for } \arh \leq a\,,
    \end{dcases}
\end{equation}
where $\Hrh \equiv H(\arh) \simeq \frac{\pi}{3} \sqrt{\frac{\gs(\Trh)}{10}}\, \frac{\Trh^2}{M_P}$ is the Hubble scale at the end of the reheating era. The BICEP/Keck bound on the tensor-to-scalar ratio implies that the Hubble parameter during inflation $H_I$ is bounded from above $H_I \leq 2.0 \times 10^{-5}~M_P$~\cite{BICEP:2021xfz}. That translates into an upper bound on the maximal temperature $\Tmax$ reached by the thermal bath during reheating given by~\cite{Bernal:2024yhu}
\begin{equation} \label{eq:Tmax}
    \Tmax \simeq \Trh \left[\frac{90}{\pi^2\, \gs}\, \frac{H_I^2\, M_P^2}{\Trh^4}\right]^\frac{\alpha}{3 (1+\omega)}.
\end{equation}

Several combinations of $\omega$ and $\alpha$ can be found in the literature; see, e.g., Refs.~\cite{Co:2020xaf, Barman:2024mqo, Bernal:2024yhu}. Particularly well-motivated cases correspond to $\omega = 0$ and $\alpha = 3/8$, which arise in the case where a massive inflaton (with an energy density that scales as nonrelativistic matter) decays with a constant decay width into SM particles~\cite{Giudice:2000ex}. Furthermore, if the inflaton energy density is diluted faster than free radiation, that is, if $\omega > 1/3$, it is not necessary for the inflaton to decay or annihilate away, and then one can have $\alpha = 1$, as in the case of kination~\cite{Spokoiny:1993kt, Ferreira:1997hj}.

%%%%%%%%%%%%%%%%%%%%%%%%%%%%%%%%%%%%%%%%%%%%%%%%%%%%%%%%%%
\section{FIMP Dark Matter} \label{sec:DM}
%%%%%%%%%%%%%%%%%%%%%%%%%%%%%%%%%%%%%%%%%%%%%%%%%%%%%%%%%%
We assume that, in the early Universe, DM is produced solely through the FIMP mechanism from annihilations of particles from the SM plasma; or, equivalently, that the FIMP production is the dominant mechanism. In that case, the evolution of the DM number density $n$ can be followed with the Boltzmann equation
\begin{equation} \label{eq:DM}
    \frac{dn}{dt} + 3\, H\, n = \sv\, n_\text{eq}^2\,,
\end{equation}
with $n_\text{eq}(T)$ being the equilibrium DM number density, and where the possibility of DM annihilation (i.e. the backreaction term) has been ignored. Additionally, $\sv(T)$ corresponds to the effective thermally-averaged cross section for DM, which is conveniently computed using the numerical implementation in the {\tt F{\small REEZE}I{\small N}} code~\cite{Bhattiprolu:2023akk, FreezeIn, Bhattiprolu:2024dmh}, which also includes the plasmon decay contribution. Note that the code only takes into account the quantum statistical effect for plasmon decay, while Maxwell-Boltzmann distributions are used for the rest of the annihilating initial states. To quantify the quantum statistical effects, we modify the code to take into account the Fermi-Dirac phase-space distribution for fermions and that of the Bose-Einstein for bosons for all the initial states (the details are in the Appendix~\ref{app:quantum_statistics}).

Taking into account the general background defined in Eqs.~\eqref{eq:T} and~\eqref{eq:H}, it is convenient to rewrite Eq.~\eqref{eq:DM} as
\begin{equation}
    \frac{d\left(n\, a^3\right)}{da} = \frac{a^2}{H}\, \sv\, n_\text{eq}^2\,.
\end{equation}
This expression is numerically solved ignoring a possible initial DM abundance from the inflationary era or from the direct decay of the inflaton.\footnote{This is typically a good assumption as long as the branching fraction Br of the inflaton into DM particles satisfies $\text{Br} \lesssim 10^{-4} \times m_\chi/(100~\text{GeV})$~\cite{Drees:2017iod, Arias:2019uol}.} To match the entire observed DM relic density it is required that
\begin{equation}
    m_\chi\, Y_0 = \frac{\Omega h^2\, \rho_c}{s_0\, h^2} \simeq 4.3 \times 10^{-10}~\text{GeV}, \label{eq:observed_DM}
\end{equation}
where $Y_0$ is the asymptotic value of the DM yield $Y(T) \equiv n(T)/s(T)$ at low temperatures, $s_0 \simeq 2.69 \times 10^3$~cm$^{-3}$ is the present entropy density~\cite{ParticleDataGroup:2024cfk}, $\rho_c \simeq 1.05 \times 10^{-5}~h^2$~GeV/cm$^3$ is the critical energy density of the universe, and $\Omega h^2 \simeq 0.12$ is the observed DM relic abundance~\cite{Planck:2018vyg}.

DM production with a low reheating temperature has been intensively studied in the literature, usually triggered by the decay of a long-lived massive particle~\cite{Giudice:2000ex, Fornengo:2002db, Pallis:2004yy, Gelmini:2006pw, Drees:2006vh, Yaguna:2011ei, Roszkowski:2014lga, Drees:2017iod, Bernal:2018ins, Bernal:2018kcw, Cosme:2020mck, Arias:2021rer, Bernal:2022wck, Bhattiprolu:2022sdd, Haque:2023yra, Chowdhury:2023jft, Ghosh:2023tyz, Becker:2023tvd, Silva-Malpartida:2023yks, Arias:2023wyg, Xu:2023lxw, Bernal:2024yhu, Banerjee:2024caa, Silva-Malpartida:2024emu}.\footnote{For studies on baryogenesis with a low reheating temperature or during an early matter-dominated phase, see Refs.~\cite{Davidson:2000dw, Giudice:2000ex, Allahverdi:2010im, Beniwal:2017eik, Allahverdi:2017edd, Nelson:2019fln,Asaka:2019ocw,Jaeckel:2022osh} and~\cite{Bernal:2017zvx, Chen:2019etb, Bernal:2022pue, Chakraborty:2022gob}, respectively. Furthermore, the production of primordial gravitational waves during reheating has recently received particular attention~\cite{Assadullahi:2009nf, Durrer:2011bi, Alabidi:2013lya, DEramo:2019tit, Bernal:2019lpc, Figueroa:2019paj, Bernal:2020ywq, Frey:2024jqy, Villa:2024jbf, Bernal:2024jim}.} In the following, the DM production after and during reheating will be carefully analyzed.

%%%%%%%%%%%%%%%%%%%%%%%%%%%%%%%%%%%%%%%%%%%%%
\subsection{After Reheating}
\begin{figure}[t!]
    \def\sepf{0.6}
    \centering
    \includegraphics[width=\sepf\columnwidth]{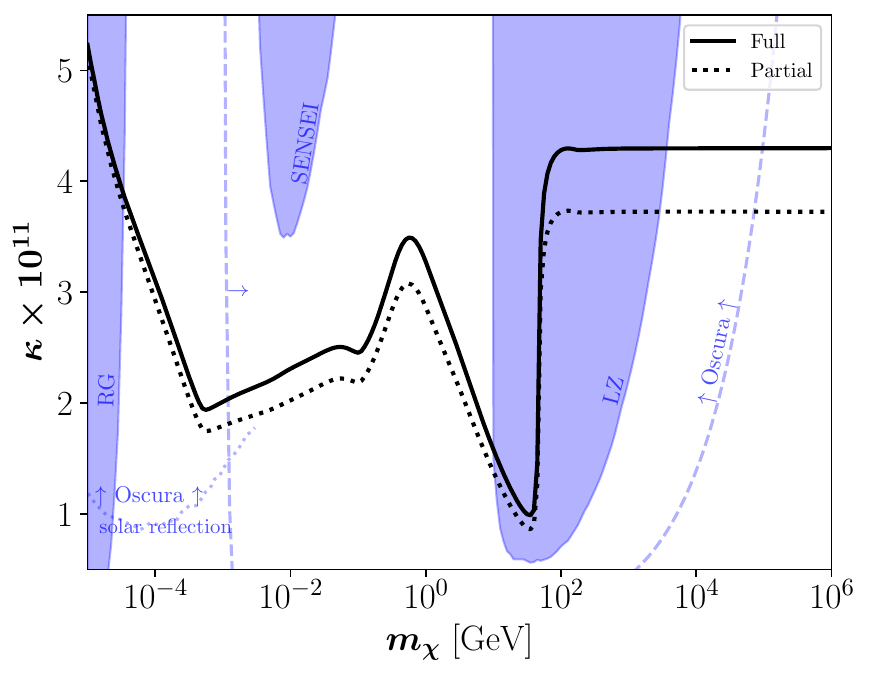}
    \caption{The required $\kappa$ as a function of DM mass $m_\chi$ in order to reproduce the observed DM relic density c.f. Eq.~\eqref{eq:observed_DM} when full quantum statistical effects are taken into account (black solid curve) and when the effects are only taken into account for plasmon decays (black dotted curve). Blue regions are already excluded by experimental measurements, while the blue dotted and dashed lines represent projected sensitivities from Oscura experiment; see the text.}
    \label{fig:high_reheating}
\end{figure}

%%%%%%%%%%%%%%%%%%%%%%%%%%%%%%%%%%%%%%%%%%%%%
In Fig.~\ref{fig:high_reheating}, we show the required value for $\kappa$ to fit the observed DM relic density, cf. Eq.~\eqref{eq:observed_DM}, as a function of $m_\chi$ for a high reheating temperature scenario $\Trh \gg m_\chi$. In general, DM is produced from the annihilation of SM particles; however, if DM is lighter than the electron or in the range $1 \lesssim m_\chi \lesssim m_Z/2$, it is produced mainly from decays of the plasmon or $Z$ bosons, respectively. Furthermore, above (below) the thick black line, the DM is overproduced (underproduced). In addition, we see that quantum statistical corrections amount to $\sim 10\%$ effects for $m_\chi \gtrsim 1$~MeV, while for $m_\chi \lesssim 1$~MeV when plasmon decays dominate, the effects decrease to around 2\% for $m_\chi \sim 10^{-2}$~MeV. In the relevant mass range, the effects are in accordance with Ref.~\cite{Heeba:2023bik} which considers a modified model where the DM is pseudo-Dirac. As we show in the following subsection, the effects due to the low reheating scenarios can be of several orders of magnitude, dwarfing the quantum statistical effects.
%%%%%%%%%%%%%%%%%%%%%%%%%%%%%%%%%%%%%%%%%%%%%%%%%%%
%%%%%%%%%%%%%%%%%%%%%%%%%%%%%%%%%%%%%%%%%% 

Because kinetic mixing allows DM to interact with SM matter, it is possible to probe the minimal freeze-in model through direct detection experiments (blue regions in Fig.~\ref{fig:high_reheating}). For $10~\text{GeV} \lesssim m_\chi \lesssim 3$~TeV, the strong limits obtained by recasting~\cite{Chu:2011be, Hambye:2018dpi} the constraints on the DM-nucleon spin-independent cross section from the LZ collaboration already rule out part of the favored parameter space~\cite{LZCollaboration:2024lux}. Furthermore, direct detection limits through electron scattering from the SENSEI collaboration are approaching the freeze-in curve for $m_\chi \sim 10^{-2}$~GeV~\cite{SENSEI:2023zdf}. For $m_\chi \lesssim 10^{-4}$~GeV, strong bounds appear from the cooling of red-giant (RG) stars due to the emission of DM~\cite{Fung:2023euv}. Finally, we note that next-generation direct detection experiments such as Oscura~\cite{Oscura:2022vmi} could have the sensitivity to probe $1~{\rm MeV}\lesssim m_\chi \lesssim 100$~TeV through electron recoil of DM in the halo (area bounded by the blue dashed lines) and $m_\chi \lesssim 1\,{\rm MeV}$ using solar-reflected DM (area above the blue dotted line)~\cite{An:2021qdl, CDEX:2023wfz, Emken:2024nox}.
 
%%%%%%%%%%%%%%%%%%%%%%%%%%%%%%%%%%%%%%%%%%%%%
\subsection{During Reheating}
\begin{figure}[t!]
    \def\sepf{0.496}
    \centering
    \includegraphics[width=\sepf\columnwidth]{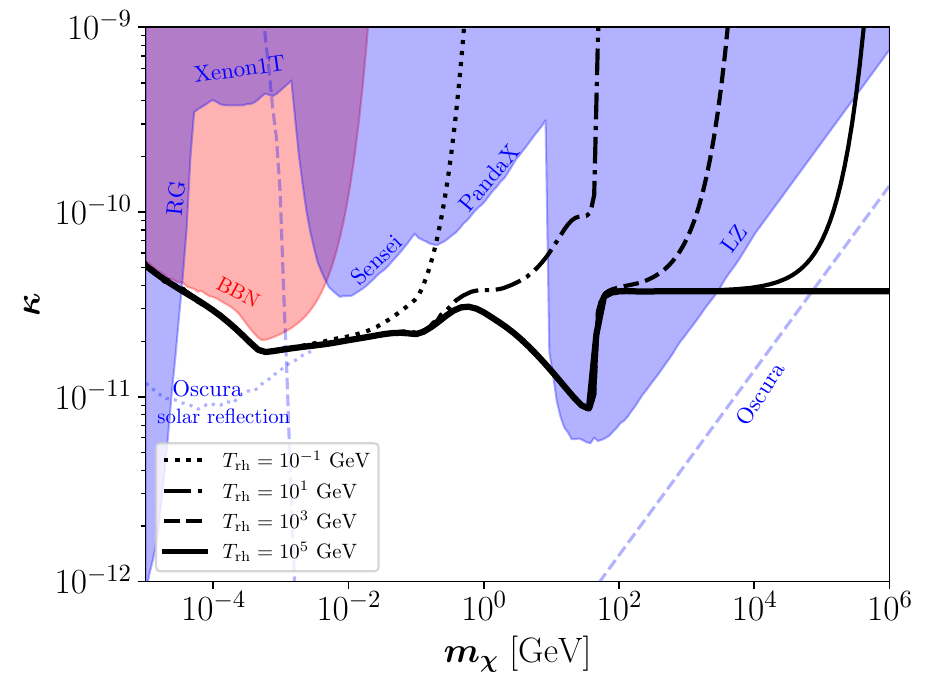}
    \includegraphics[width=\sepf\columnwidth]{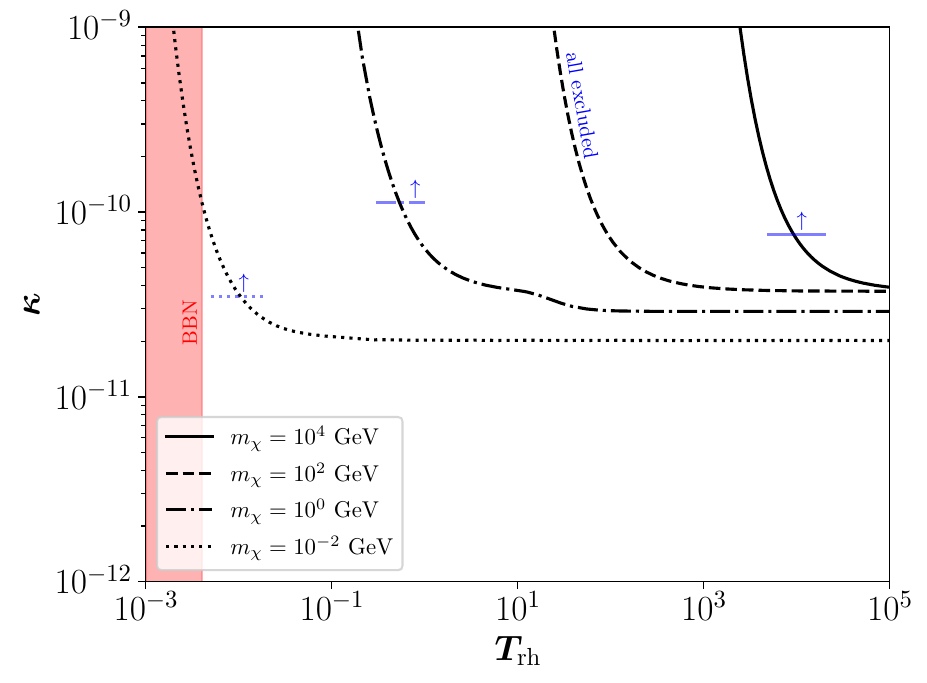}
    \caption{Left panel: Parameter space required to match the whole observed DM abundance for a large-temperature reheating (thick black line), or low-temperature reheating $\Trh$ (thin black lines of different styles), for an {\it instantaneous} reheating. The blue areas are in tension with experimental data, while the blue dotted and dashed lines correspond to projections of next-generation direct detection experiments, see text. Right panel: the required values of $\kappa$ for a given $T_{\rm rh}$ are shown, for four representative values of $m_\chi$. The horizontal lines are current experimental constraints where the values of $\kappa$ above the lines are excluded. In both panels, the red areas indicate $\Trh < \Tbbn$.}
    \label{fig:EMD0}
\end{figure}

%%%%%%%%%%%%%%%%%%%%%%%%%%%%%%%%%%%%%%%%%%%%%
The thick black line in the left panel of Fig.~\ref{fig:EMD0} shows the parameter space that fits the entire observed abundance of DM, assuming a large reheating temperature: this corresponds to the case where $\Trh \gg m_\chi$ and therefore the bulk of the DM is produced during the radiation-dominated era; cf. Fig.~\ref{fig:high_reheating}. However, $\Trh$ could be comparable or even much smaller than the DM mass. The four thinner black lines of different styles in the left panel of Fig.~\ref{fig:EMD0} show the parameter space where the entire abundance of DM is fitted, for different low-temperature reheating scenarios, for an {\it instantaneous} reheating; that is, $\Tmax = \Trh$ which are in agreement with Ref.~\cite{Boddy:2024vgt}. In this case, the production of DM is exponentially reduced, since only SM particles with velocities in the tail of the distribution have sufficient kinetic energy to produce DM states, which must be compensated for with an exponential increase in the portal coupling $\kappa$.\footnote{The exponential increase in the portal coupling is very generic and can be found in different scenarios, e.g. Refs.~\cite{Cosme:2023xpa, Cosme:2024ndc, Arcadi:2024wwg, Arcadi:2024obp, Lebedev:2024vor}.} In the red area, $\Trh < \Tbbn$ is required, and therefore low reheating cannot be used to enhance $\kappa$ in this region.
%%%%%%%%%%%%%%%%%%%%%%%%%%%%%%%%%%%%%%%%%%%%%%%%%%%
%%%%%%%%%%%%%%%%%%%%%%%%%%%%%%%%%%%%%%%%%% 

The blue area on the left panel of Fig.~\ref{fig:EMD0} is excluded by different experimental results: for $10^1~\text{GeV} \lesssim m_\chi$ direct searches from DM scattering off of nuclei by LZ~\cite{LZCollaboration:2024lux}, while $10^{-3}~\text{GeV}\lesssim m_\chi \lesssim 10^{-1}$~GeV and $10^{-1}~\text{GeV}\lesssim m_\chi \lesssim 10^1$~GeV by SENSEI~\cite{SENSEI:2023zdf} and PandaX~\cite{PandaX:2022xqx}, respectively, from DM-electron scatterings. For $10^{-4}~\text{GeV}\lesssim m_\chi \lesssim 10^{-3}$~GeV, $\kappa \gtrsim 4\times 10^{-10}$ due to the reinterpretation of XENON1T data~\cite{XENON:2019gfn, XENON:2020rca} using the reflection of DM by electrons in the Sun~\cite{An:2021qdl}. Interestingly, LZ already rules out the range of the parameter space corresponding to $10~\text{GeV} \lesssim m_\chi \lesssim 3$~TeV, for the case of high-temperature reheating~\cite{Chu:2011be, Hambye:2018dpi}. The exclusion becomes even stronger if the reheating temperature is low, due to the required increase of the portal coupling. In fact, large regions of the parameter space favored by low-temperature reheating are already excluded. Remarkably, the projected sensitivity to $\chi$-electron scattering~\cite{Oscura:2022vmi, Emken:2024nox} of the Oscura experiment~\cite{Oscura:2022vmi} will probe most of the parameter space considered in this work, that is, $10^{-5}\lesssim m_\chi/{\rm GeV} \lesssim 10^{5}$ (blue dotted and dashed lines). The right panel of Fig.~\ref{fig:EMD0} emphasizes the required increase of $\kappa$ in the case of low $\Trh$, for instantaneous reheating and different masses of DM. Again, the red area is in tension with BBN.

%%%%%%%%%%%%%%%%%%%%%%%%%%%%%%%%%%%%%%%%%%%%%%%%%%%
\begin{figure}[t!]
    \def\sepf{0.496}
    \centering
    \includegraphics[width=\sepf\columnwidth]{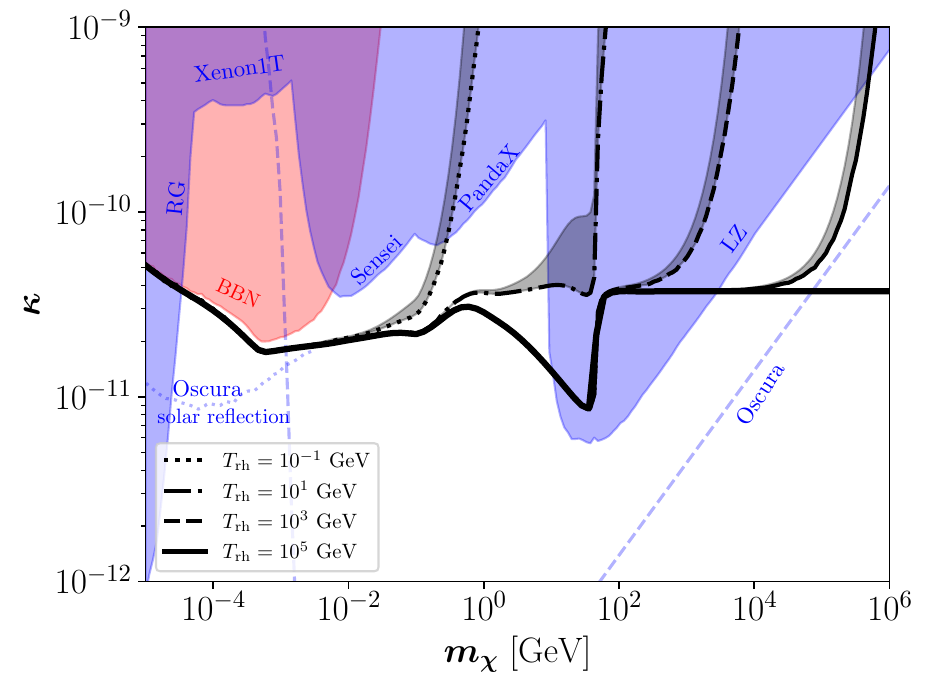}
    \includegraphics[width=\sepf\columnwidth]{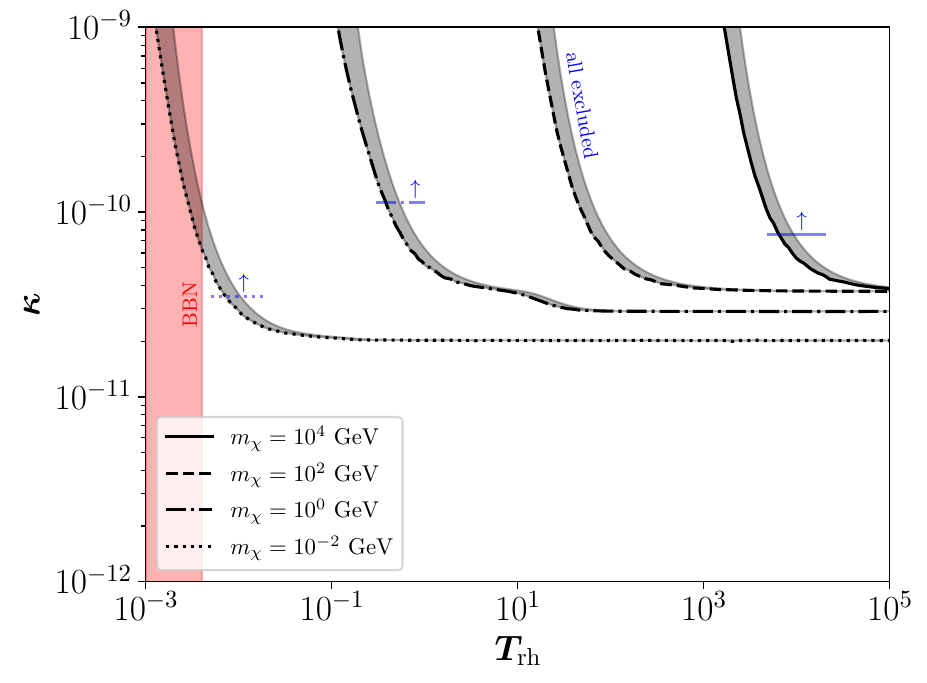}
    \caption{Parameter space required to match the whole observed DM abundance for a large-temperature reheating (thick black line), or low-temperature reheating (thin gray bands), for $\omega = 0$, $\alpha = 3/8$. The upper (lower) bound of the bands correspond to the case where the duration of reheating in minimized (maximized). The blue areas are in tension with experimental data; while the blue lines correspond to projections of next-generation direct detection experiments. In the red areas $\Trh < \Tbbn$.}
    \label{fig:EMD}
\end{figure} 
%%%%%%%%%%%%%%%%%%%%%%%%%%%%%%%%%%%%%%%%%% 
We remind the reader that Fig.~\ref{fig:EMD0} corresponds to a case where reheating occurs instantaneously. However, this is not expected to be the case. Reheating, as any physical process, may be fast but last a finite amount of time. As explained in Section~\ref{sec:framework}, in the case of a non-instantaneous reheating, apart from $\Trh$, the values for $\Tmax$, $\omega$, and $\alpha$ must be fixed to determine the evolution of the background. In that context, the thin black lines (corresponding to the lower bound of the gray bands) of Fig.~\ref{fig:EMD} show the parameter space compatible with the total DM relic abundance, for $\omega = 0$ and $\alpha = 3/8$. A {\it maximal} duration of reheating was assumed, by maximizing the value of $\Tmax$; cf. Eq.~\eqref{eq:Tmax}. The upper bound of the bands corresponds to the case where reheating is instantaneous; see Fig.~\ref{fig:EMD0}. Therefore, the thickness of the bands brackets the uncertainty of the duration of the reheating era, for a fixed $\Trh$. The bands are generally narrow because of the large injection of entropy from the inflaton decay into SM particles. We remind the reader that this case with $\omega = 0$ and $\alpha = 3/8$ corresponds to a massive inflaton that decays into SM states, diluting the DM population produced during reheating. The results for an instantaneous reheating (meaning no reheating era at all) and a reheating with a large entropy injection are not very different, as seen in Fig.~\ref{fig:EMD}.

%%%%%%%%%%%%%%%%%%%%%%%%%%%%%%%%%%%%%%%%%%%%%%%%%%%
\begin{figure}[t!]
    \def\sepf{0.496}
    \centering
    \includegraphics[width=\sepf\columnwidth]{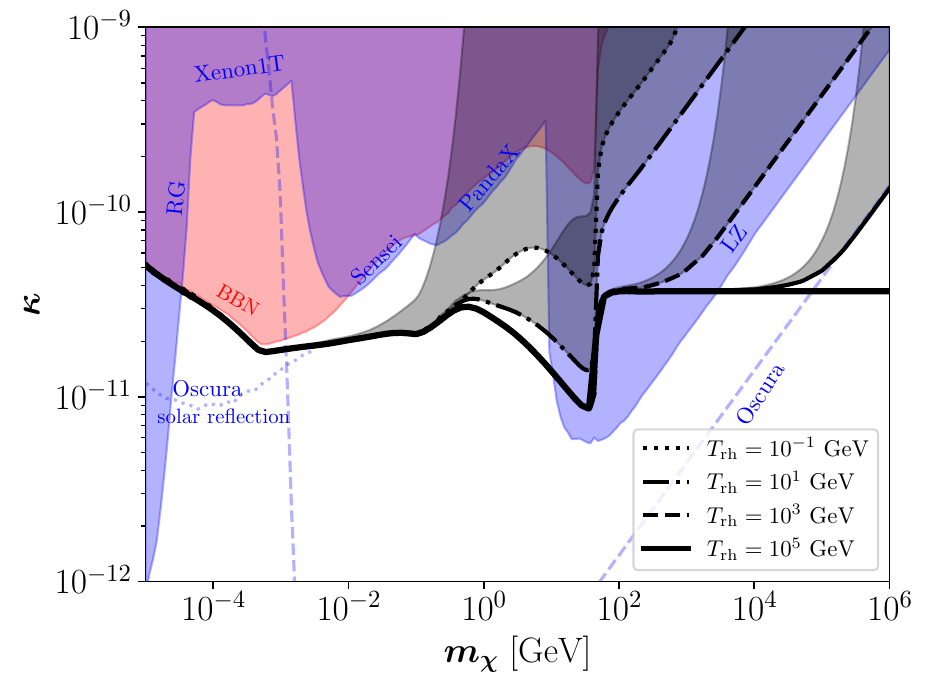}
    \includegraphics[width=\sepf\columnwidth]{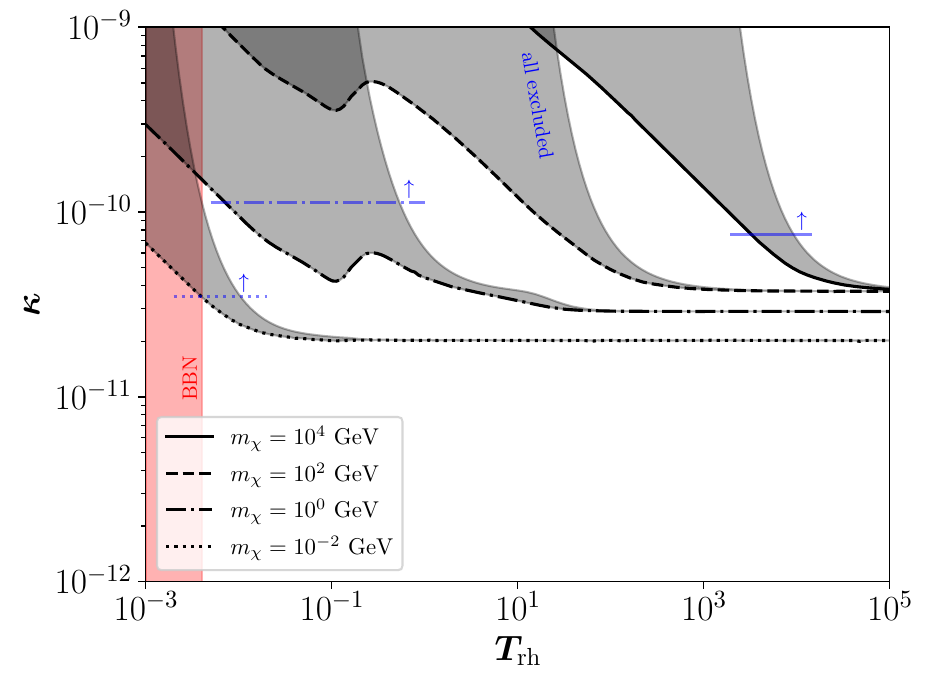}
    \caption{Same as Fig.~\ref{fig:EMD}, but for a kination scenario with $\omega = \alpha = 1$.}
    \label{fig:kination}
\end{figure} 
%%%%%%%%%%%%%%%%%%%%%%%%%%%%%%%%%%%%%%%%%% 
The situation can change dramatically in a case without injection of entropy. For example, Fig.~\ref{fig:kination} compares to Fig.~\ref{fig:EMD} but for $\omega = \alpha = 1$ and corresponds to a kination-like period. As in this scenario the SM entropy is conserved ($\alpha = 1$), during the reheating era, the only impact on the DM abundance comes from the increase on the Hubble parameter. This must be compensated for with a slight increase in portal coupling $\kappa$, compared to the case of high-temperature reheating.

Before closing, we note that the self-interactions of the DM could also constrain the model~\cite{Feng:2009hw}; however, they bound the dark gauge coupling $e'$ and not directly $\kappa$. To have a sense of the bound on $e'$, let us consider the cross section of self-interaction of DM through a very light dark photon from the dominant $t$-channel contribution~\cite{Feng:2009hw}
\begin{equation}
    \sigma_{\chi\chi} \simeq \frac{e'{}^{4}}{4\pi\, m_\chi^{2}\, v^{4}}\ln\frac{m_\chi^{2}\, v^{2}}{m_{\gamma'}^{2}}\,.
\end{equation}
Using the bound $\sigma_{\chi\chi}/m_\chi\lesssim1\,{\rm cm}^{2}/{\rm g}$~\cite{Kaplinghat:2015aga}, we have
\begin{equation}
    e' \lesssim 5\times10^{-3}\left(\frac{m_\chi}{1\,{\rm GeV}}\right)^{3/4}\left(\frac{v}{10^{-3}}\right)\left[\frac{\ln\left(m_\chi^{2}v^{2}/m_{\gamma'}^{2}\right)}{100}\right]^{-1/4}.
\end{equation}
In addition, the requirement to avoid chemical equilibrium between the dark and visible sectors implies $\kappa \lesssim 10^{-7}$, and is therefore always satisfied in the parameter space of interest.

%%%%%%%%%%%%%%%%%%%%%%%%%%%%%%%%%%%%%%%%%%%%%%%%%%%%%%%%%%
\section{Conclusions} \label{sec:concl}
%%%%%%%%%%%%%%%%%%%%%%%%%%%%%%%%%%%%%%%%%%%%%%%%%%%%%%%%%%
It is commonly assumed that the cosmic inflationary era is followed by an {\it instantaneous} reheating, driven by the sudden decay of the inflaton into standard-model (SM) particles. Although this assumption may be inconsequential for high reheating temperatures $\Trh$, it becomes critical once one allows $\Trh$ to be around or even smaller than the scale at which DM is produced in the early universe. In that case, details of the reheating era, such as the equation-of-state parameter of the inflaton and the scaling of the SM temperature, attain paramount importance.

In this work, we have explored the impact of a {\it non-instantaneous} reheating phase on the parameter space required to fit the entire observed DM abundance, in the context of the minimal freeze-in DM scenario. After revisiting the case of an instantaneous reheating, we considered a reheating scenario where the inflaton scales as non-relativistic matter and decays with a constant decay width to SM particles, and a kination-like case. We place particular emphasis on bracketing the uncertainty in relation to the duration of the reheating phase and on the effect of quantum statistics when computing the DM production rate. A detailed understanding of the parameter of the model is particularly pertinent because the extensive experimental program for DM detection is already probing the parameter space that yields the correct DM abundance. We have explored a wide mass range of DM from $10^{-2}$~MeV to $10^3$~TeV where the region with $m_\chi < 3\times 10^{-2}$~MeV is already excluded by cooling of red-giant stars, while for higher-mass regions, some are in tension with current direct detection experiments with bounds that are sensitive to low-reheating scenarios $\Trh \lesssim m_\chi$. Interestingly, next-generation experiments such as SENSEI~\cite{Tiffenberg:2017aac}, DAMIC-M~\cite{Castello-Mor:2020jhd} and Oscura~\cite{Oscura:2022vmi} will be able to probe entirely $m_\chi \lesssim 10^2$~TeV, and the reach of the mass range can be several orders of magnitude higher for low reheating scenarios. Finally, in the case of a detection, the characterization of the reheating era will require the complementary information from other messengers such as primordial gravitational waves.

%%%%%%%%%%%%%%%%%%%%%%%%%%%%%%%%%%%%%%%%%%%
\acknowledgments
%%%%%%%%%%%%%%%%%%%%%%%%%%%%%%%%%%%%%%%%%%%
NB received funding from the Grant PID2023-151418NB-I00 funded by MCIU/AEI/10.13039 /501100011033/ FEDER, UE, and acknowledges support by Institut Pascal at Université Paris-Saclay during the Paris-Saclay Astroparticle Symposium 2024, with the support of the P2IO Laboratory of Excellence (program ``Investissements d’avenir'' ANR-11-IDEX-0003-01 Paris-Saclay and ANR-10-LABX-0038), the P2I axis of the Graduate School of Physics of Université Paris-Saclay, as well as IJCLab, CEA, IAS, OSUPS, the IN2P3 master project UCMN, and APPEC. CSF acknowledges the support by Fundacão de Amparo à Pesquisa do Estado de São Paulo (FAPESP) Contracts No. 2019/11197-6 and 2022/00404-3 and Conselho Nacional de Desenvolvimento Científico e Tecnológico (CNPq) under Contract No. 304917/2023-0. OZ has been partially supported by Sostenibilidad-UdeA, the UdeA/CODI Grants 2022-52380 and 2023-59130, and the Ministerio de Ciencias Grant CD 82315 CT ICETEX 2021-1080. We thank the organizers of the XV Latin American Symposium on High Energy Physics (SILAFAE) for providing a welcoming environment that facilitated valuable discussions relevant to this work.

%%%%%%%%%%%%%%%%%%%%%%%%%%%%%%%%%%%%%%%
\appendix
%%%%%%%%%%%%%%%%%%%%%%%%%%%%%%%%%%%%%%%%%%
%%%%%%%%%%%%%%%%%%%%%%%%%%%%%%%%%%%%%%%%
\section{Quantum statistical effects} \label{app:quantum_statistics}
%%%%%%%%%%%%%%%%%%%%%%%%%%%%%%%%%%%%%%%%
In this appendix, we review how quantum statistical effects can be taken into account. For freeze-in, they were previously considered in {\tt micrOMEGAs}~\cite{Belanger:2001fz, Alguero:2023zol}. 

The collision term for freeze-in of $ab\to\chi\overline{\chi}$ where $\chi \overline{\chi}$ is a pair DM particle-antiparticle while $a$ and $b$ are assumed to be in thermal equilibrium is
\begin{align}
    C\left(ab\leftrightarrow\chi\overline{\chi}\right) &= \Lambda_{ab}^{\chi\overline{\chi}}\left|{\cal M}\left(ab\to\chi\overline{\chi}\right)\right|^{2}\nonumber \\
    &\quad \times \left[f_{a}^{{\rm eq}}f_{b}^{{\rm eq}}\left(1+\eta_{\chi}f_{\chi}\right)\left(1+\eta_{\chi}f_{\overline{\chi}}\right)-f_{\chi}f_{\overline{\chi}}\left(1+\eta_{a}f_{a}^{{\rm eq}}\right)\left(1+\eta_{b}f_{b}^{{\rm eq}}\right)\right],
\end{align}
where $\eta_{i}=+1\left(-1\right)$ for boson (fermion), $f_{i}^{{\rm eq}}=\left(e^{E_{i}/T}-\eta_{i}\right)^{-1}$ and we have assumed CP conservation in the squared amplitude and defined
\begin{equation}
    \Lambda_{ab}^{\chi\overline{\chi}} \equiv \int d\Pi_{\chi}d\Pi_{\overline{\chi}}d\Pi_{a}d\Pi_{b}\left(2\pi\right)^{4}\delta^{\left(4\right)}\left(p_{a}+p_{b}-p_{\chi}-p_{\overline{\chi}}\right),\quad d\Pi_{i}\equiv\frac{d^{3}p_{i}}{\left(2\pi\right)^{3}2E_{i}}\,.
\end{equation}
Notice that in the squared amplitude we have sum (not averaged) over final and initial degrees of freedom.

For freeze-in, as $f_{\chi}$ and $f_{\overline{\chi}} \ll 1$ at all time, one can drop them to get\footnote{Refs.~\cite{Yin:2023jjj, Sakurai:2024apm} show that for boson $\chi$ and $\bar\chi$, it is possible that $f_{\chi}, f_{\overline{\chi}} \gg 1$ due to population of low momentum modes leading to Bose enhancement in their production. For the low-reheating scenarios under consideration, this effect is negligible.} 
\begin{equation}
    C\left(ab\leftrightarrow\chi\overline{\chi}\right) = \Lambda_{ab}^{\chi\overline{\chi}}\left|{\cal M}\left(ab\to\chi\overline{\chi}\right)\right|^{2}f_{a}^{{\rm eq}}f_{b}^{{\rm eq}}\,,
\end{equation}
and in this case, the spin of freeze-in DM is not relevant. 
Using the definition of cross section (summing over both final and initial degrees of freedom)
\begin{equation}
    \sigma_{ab\to\chi\overline{\chi}}\left(s\right) \equiv \frac{1}{2\, s\, \beta\left(1,\frac{m_{a}^{2}}{s},\frac{m_{b}^{2}}{s}\right)}\int d\Phi_{2}\left|{\cal M}\left(ab\to\chi\overline{\chi}\right)\right|^{2},
\end{equation}
where $d\Phi_{2} \equiv \left(2\pi\right)^{4} \delta^{\left(4\right)}\left(p_{a}+p_{b}-p_{\chi}-p_{\overline{\chi}}\right) d\Pi_{\chi} d\Pi_{\overline{\chi}}$ is the two-body phase space element, we have
\begin{equation}
    C\left(ab\leftrightarrow\chi\overline{\chi}\right) = \int d\Pi_{a} d\Pi_{b}\, 2\, s\, \beta\left(1, \frac{m_{a}^{2}}{s}, \frac{m_{b}^{2}}{s}\right) \sigma_{ab\to\chi\overline{\chi}}\left(s\right) f_{a}^{{\rm eq}}\, f_{b}^{{\rm eq}}\,,
\end{equation}
with $\beta\left(1, v, w\right) \equiv \sqrt{\left(1 - v - w\right)^{2} - 4 v\, w}$. Inserting the identity $1 = \int d^{4}P\, \delta^{(4)}(P-p_{a}-p_{b})$, we have to carry out the integral and using the identity for two-body phase space integral
\begin{equation}
    \frac{1}{8\pi}\beta\left(1,\frac{m_{a}^{2}}{s},\frac{m_{b}^{2}}{s}\right) = \int d\Pi_{a}d\Pi_{b}\left(2\pi\right)^{4}\delta^{\left(4\right)}\left(P-p_{a}-p_{b}\right),
\end{equation}
where with $s=P^{2}$, we have
\begin{equation}
    C\left(ab\leftrightarrow\chi\overline{\chi}\right) = \frac{1}{8\pi}\int\frac{d^{4}P}{\left(2\pi\right)^{4}}\, 2s\, \beta^2\left(1,\frac{m_{a}^{2}}{s},\frac{m_{b}^{2}}{s}\right)\, \sigma_{ab\to\chi\overline{\chi}}\left(s\right)\, f_{a}^{{\rm eq}}\, f_{b}^{{\rm eq}}\,.
\end{equation}

In the rest frame of the two-body scattering $\overline{P}=\left(\sqrt{s}, 0, 0, 0\right)$, we have
\begin{align}
    \overline{E}_{a} &= \sqrt{\overline{p}_{ab}^{2}+m_{a}^{2}}=\frac{\sqrt{s}}{2}\left(1+\frac{m_{a}^{2}}{s}-\frac{m_{b}^{2}}{s}\right)\equiv\frac{\sqrt{s}}{2}\, \epsilon_{a} \,,\\
    \overline{E}_{b} &= \sqrt{\overline{p}_{ab}^{2}+m_{b}^{2}}=\frac{\sqrt{s}}{2}\left(1-\frac{m_{a}^{2}}{s}+\frac{m_{b}^{2}}{s}\right)\equiv\frac{\sqrt{s}}{2}\, \epsilon_{b} \,
\end{align}
where
\begin{equation}
    \overline{p}_{ab} = \frac{\sqrt{s}}{2}\, \beta\left(1,\frac{m_{a}^{2}}{s},\frac{m_{b}^{2}}{s}\right) \equiv \frac{\sqrt{s}}{2}\, \beta_{ab}\,.
\end{equation}
Notice that $E_{a}$ and $E_{b}$ are given in the comoving frame (of the thermal bath) $P=\left(P_{0},\vec{P}\right)$ which are related to $\overline{E}_{a}$, $\overline{E}_{b}$ and $\overline{p}$ through a
Lorentz boost
\begin{equation}
    E_{a} = \gamma \overline{E}_{a} + \gamma\, \beta\, \overline{p}_{ab}\, c_{\theta}\,, \qquad E_{b} = \gamma\, \overline{E}_{b} - \gamma\, \beta\, \overline{p}_{ab}\, c_{\theta}\,,
\end{equation}
where $c_{\theta} \equiv \cos\theta$ with $\theta$ the angle between momenta of $a$ and $b$ to the boost direction and
\begin{equation}
    \gamma = \frac{P_{0}}{\sqrt{s}}\,, \qquad \gamma\, \beta = \frac{|\vec{P}|}{\sqrt{s}} = \frac{\sqrt{P_{0}^{2} - s}}{\sqrt{s}}\,.
\end{equation}
Next we can rewrite
\begin{equation}
    d^{4}P = 2 \pi\, |\vec{P}|^{2}\, d|\vec{P}|\, dP_{0}\, dc_{\theta} = \pi \sqrt{P_{0}^{2}-s}\, ds\, dP_{0}\, dc_{\theta}\,,
\end{equation}
where $-1\leq c_{\theta}\leq1$, $\sqrt{s}\leq P_{0}<\infty$ and
$s_{{\rm min}} \equiv \min\left[\left(m_{a} + m_{b}\right)^{2},\, 4 m_{\chi}^{2}\right] \leq s< \infty$. The collision term becomes
\begin{equation}
    C\left(ab\leftrightarrow\chi\overline{\chi}\right) = \int \frac{ds\, dP_{0}\, dc_{\theta}}{(2\pi)^{4}}\, \sqrt{P_{0}^{2} - s}\, \overline{p}_{ab}^{2}\, \sigma_{a b \to \chi \overline{\chi}} (s)\, f_{a}^{{\rm eq}}\, f_{b}^{{\rm eq}}\,.
\end{equation}
Integrating over $c_{\theta}$, we have
\begin{align}
    e^{x y} \int_{-1}^{1}\frac{dc_{\theta}}{2}\, f_{a}^{{\rm eq}}\, f_{b}^{{\rm eq}} &= \frac{1 + \frac{1}{\sqrt{x^{2}-1}\, \beta_{ab}\, y}}{1 - \eta_{a}\, \eta_{b}\, e^{-x\, y}}\, \ln\frac{\left(1 - \eta_{a}\, e^{-x_{a+} y}\right)\left(1 - \eta_{b}\, e^{-x_{b+} y}\right)}{\left(1 - \eta_{a}\, e^{-x_{a-} y}\right)\left(1 - \eta_{b}\, e^{-x_{b-} y}\right)}\nonumber\\
    &\equiv S\left(x,y,r_{a},r_{b},\eta_{a},\eta_{b}\right),
\end{align}
where we have defined $x \equiv P_{0}/\sqrt{s}$, $y \equiv \sqrt{s}/T$,
$r_{i} \equiv m_{i}/\sqrt{s}$ and
\begin{equation}
    x_{a\pm} \equiv \frac{1}{2} \left(x\, \epsilon_{a} \pm \sqrt{x^{2} - 1}\, \beta_{ab}\right), \qquad x_{b\pm} \equiv \frac{1}{2} \left(x\, \epsilon_{b} \pm \sqrt{x^{2} - 1}\, \beta_{ab}\right).
\end{equation}
Finally, the collision term becomes
\begin{equation}
    C\left(ab\leftrightarrow\chi\overline{\chi}\right) = \frac{T}{8\pi^{4}} \int_{s_{{\rm min}}}^{\infty} ds\, \sqrt{s}\, \overline{p}_{ab}^{2}\, \sigma_{a b \to \chi \overline{\chi}}(s)\, \tilde{{\cal K}}_{1}\, ,
\end{equation}
where we have defined
\begin{equation}
    \tilde{{\cal K}}_{1}\left(y, r_{a}, r_{b}, \eta_{a}, \eta_{b}\right) \equiv y \int_{1}^{\infty} dx\, \sqrt{x^{2} - 1}\, e^{- x y}\, S\left(x,y,r_{a},r_{b},\eta_{a},\eta_{b}\right).
\end{equation}
Setting $\eta_{a} = \eta_{b} = 0$, and $S(x, y, r_{a}, r_{b}, 0, 0) = 1$, one recovers the modified Bessel function of second kind 
\begin{equation}
    {\cal K}_{1}\left(y\right) = \tilde{{\cal K}}_{1}\left(y, r_{a}, r_{b}, 0, 0\right).
\end{equation}

In the freeze-in model considered in this work, as far as the masses and statistics are concerned, we have $a=b$ and $\chi=\overline{\chi}$ which lead to
\begin{align}
    \tilde{{\cal K}}_{1}\left(y, r_{a}, \eta_{a}\right) &= y \int_{1}^{\infty} dx\, \sqrt{x^{2} - 1}\, e^{-x y}\, S\left(x, y, r_{a}, \eta_{a}\right),\\
    S\left(x, y, r_{a}, \eta_{a}\right) &= \frac{1 + \frac{2}{\sqrt{x^{2} - 1}\, \beta_{a}\, y}\, \ln\frac{\left(1 - \eta_{a}\, e^{-x_{+} y}\right)}{\left(1 - \eta_{a}\, e^{-x_{-} y}\right)}}{1 - \eta_{a}^{2}\, e^{-x y}}\,,
\end{align}
with
\begin{equation}
    x_{\pm} = \frac{1}{2} \left(x \pm \sqrt{x^{2} - 1}\, \beta_{a}\right), \qquad \beta_{a} = \sqrt{1 - 4\, r_{a}^{2}}\,.
\end{equation}
The relevant fermionic initial states are lepton-antilepton and quark-antiquark pairs $\ell \overline\ell$, $q \overline q$ while the bosonic initial states are $\pi^+ \pi^-$, $K^+ K^-$, and $W^+ W^-$.

%%%%%%%%%%%%%%%%%%%%%%%%%
\bibliographystyle{JHEP}
\bibliography{biblio}
%%%%%%%%%%%%%%%%%%%%%%%%%
\end{document}